\documentclass[twocolumn]{aastex63}

\usepackage{hyperref}
\usepackage{rotating}

\shortauthors{Foo et al.}
\graphicspath{{./}{figures/}}

\begin{document}

\title{The Koi Pond: A Strongly Lensed Protocluster Core hosting a Diverse Population of DSFGs}


\correspondingauthor{Nicholas Foo}
\email{nfoo1@asu.edu}

\author[0000-0002-7460-8460]{Nicholas Foo}
\affiliation{School of Earth \& Space Exploration, Arizona State University, Tempe, AZ 85287-1404, USA}

\author[0000-0001-5429-5762]{Kevin C. Harrington}
\affiliation{Joint ALMA Observatory, Alonso de C{\'o}rdova 3107, Vitacura, Casilla 19001, Santiago de Chile, Chile}
\affiliation{National Astronomical Observatory of Japan,
Los Abedules 3085 Oficina 701, Vitacura 763 0414, Santiago, Chile}
\affiliation{European Southern Observatory, Alonso de C{\'o}rdova 3107, Vitacura, Casilla 19001, Santiago de Chile, Chile}
\affiliation{Instituto de Estudios Astrofísicos, Facultad de Ingeniería y 455 Ciencias, Universidad Diego Portales, Av. Ejército Libertador 441, Santiago, Chile}

\author[0000-0003-1625-8009]{Brenda L. Frye}
\affiliation{Department of Astronomy/Steward Observatory, University of Arizona, 933 N Cherry Ave, Tucson, AZ, 85721-0009}

\author[0000-0001-9394-6732]{Patrick S. Kamieneski}
\affiliation{School of Earth \& Space Exploration, Arizona State University, Tempe, AZ 85287-1404, USA}
\affiliation{Department of Physics and Astronomy, Chalmers University of Technology, SE-412 96 Gothenburg, Sweden}

\author[0000-0002-1173-2579]{Melanie Kaasinen}
\affiliation{European Southern Observatory, Karl-Schwarzschild-Strasse 2, D-85748 Garching, Germany}

\author[0000-0002-6150-833X]{Rafael Ortiz III} 
\affiliation{School of Earth \& Space Exploration, Arizona State University, Tempe, AZ 85287-1404, USA}

\author[0000-0001-9369-6921]{Alex Pigarelli}
\affiliation{School of Earth \& Space Exploration, Arizona State University,
Tempe, AZ 85287-1404, USA}

\author[0009-0007-0782-0721]{Gibson B.\ Bowling}
\affiliation{School of Earth \& Space Exploration, Arizona State University,
Tempe, AZ 85287-1404, USA}


\author[0000-0002-4140-0428]{Bel\'{e}n Alcalde Pampliega}
\affiliation{European Southern Observatory, Alonso de C{\'o}rdova 3107, Vitacura, Casilla 19001, Santiago de Chile, Chile}

\author[0009-0007-8574-5720]{Joe Bhangal}
\affiliation{Department of Physics \& Astronomy, University of British Columbia, 6224 Agricultural Road, Vancouver, BC V6T 1Z1, Canada}

\author[0000-0001-6650-2853]{Timothy Carleton}
\affiliation{School of Earth \& Space Exploration, Arizona State University,
Tempe, AZ 85287-1404, USA}

\author[0000-0003-3921-3313]{Jianhang Chen}
\affiliation{Max-Planck-Institut für extraterrestrische Physik, Giessenbachstrasse, D-85748 Garching, Germany}

\author[0000-0003-3329-1337]{Seth H. Cohen}
\affiliation{School of Earth \& Space Exploration, Arizona State University, Tempe, AZ 85287-1404, USA}

\author[0000-0002-7288-2925]{Camila de Sá-Freitas}
\affiliation{European Southern Observatory, Alonso de C´ordova 3107, Vitacura, Casilla 19001, Santiago de Chile, Chile}

\author[0000-0001-9065-3926]{Jose Diego}
\affiliation{FCA, Instituto de Fisica de Cantabria (UC-CSIC), Av.  de Los Castros s/n, E-39005 Santander, Spain}

\author[0000-0002-7714-688X]{Román Fernández Aranda}
\affiliation{Centro de Astrobiología (CAB), CSIC-INTA, Carretera de Ajalvir km 4, Torrejón de Ardoz 28850, Madrid, Spain}

\author[0000-0002-4223-2016]{Carlos Garcia Diaz}
\affiliation{Department of Astronomy, University of Massachusetts, Amherst, MA 01003, USA}

\author[0000-0003-3418-2482]{Nikhil Garuda}
\affiliation{Department of Astronomy/Steward Observatory, University of Arizona, 933 N Cherry Ave, Tucson, AZ, 85721-0009}

\author[0000-0002-2640-5917]{Eric F. Jim\'{e}nez-Andrade}
\affiliation{Instituto de Radioastronom\'{i}a y Astrof\'{i}sica, Universidad Nacional Aut\'{o}noma de M\'{e}xico, Antigua Carretera a P\'{a}tzcuaro \# 8701, Ex-Hda. San Jos\'{e} de la Huerta, Morelia, Michoac\'{a}n, C.P. 58089, M\'{e}xico}

\author[0000-0001-9773-7479]{Daizhong Liu}
\affiliation{Purple Mountain Observatory, Chinese Academy of Sciences, 10
Yuanhua Road, Nanjing 210023, China}

\author[0000-0001-9969-3115]{James D. Lowenthal}
\affiliation{Smith College, Northampton, MA 01063, USA}

\author[0000-0003-2475-124X]{Allison Man}
\affiliation{Department of Physics \& Astronomy, University of British Columbia, 6224 Agricultural Road, Vancouver, BC V6T 1Z1, Canada}

\author[0000-0003-1832-4137]{Allison Noble}
\affiliation{School of Earth \& Space Exploration, Arizona State University, Tempe, AZ 85287-1404, USA}

\author[0000-0002-2282-8795]{Massimo Pascale}
\affiliation{Department of Astronomy, University of California, 501 Campbell Hall \#3411, Berkeley, CA 94720, USA}

\author[0000-0001-9705-2461]{Francesca Rizzo}
\affiliation{Kapteyn Astronomical Institute, University of Groningen, Landleven 12, 9747 AD, Groningen, The Netherlands}

\author[0000-0002-8999-9636]{Hannah R. Stacey}
\affiliation{European Southern Observatory (ESO), Karl-Schwarzschild Str. 2, D-85748 Garching bei München, Germany}
\affiliation{Max Planck Institute for Astrophysics, Karl-Schwarzschild Str. 1, D-85748 Garching bei München, Germany}

\author[0000-0002-4444-8929]{Amit Vishwas}
\affiliation{Cornell Center for Astrophysics and Planetary Science, Cornell University, Space Sciences Building, Ithaca, NY 14853, USA}

\author[0000-0002-9279-4041]{Q. Daniel Wang}
\affiliation{Department of Astronomy, University of Massachusetts, Amherst, MA 01003, USA}

\author[0000-0001-8156-6281]{Rogier A. Windhorst}
\affiliation{School of Earth \& Space Exploration, Arizona State University, Tempe, AZ 85287-1404, USA}

\author[0000-0001-9163-0064]{Ilsang Yoon}
\affiliation{National Radio Astronomy Observatory, 520 Edgemont Road, Charlottesville, VA 22903}

\author[0000-0001-7095-7543]{Min S. Yun}
\affiliation{Department of Astronomy, University of Massachusetts, Amherst, MA 01003, USA}

\author[0000-0002-6922-469X]{Dazhi Zhou}
\affiliation{Department of Physics and Astronomy, University of British Columbia, 6225 Agricultural Rd., Vancouver, V6T 1Z1, Canada}



\begin{abstract}
We present James Webb Space Telescope (JWST) and Atacama Large Millimeter Array (ALMA) observations of PJ0846+15, \textit{The Koi Pond}, a strongly lensed protocluster core at Cosmic Noon. This field offers a magnified view of 11 dusty star-forming galaxies (DSFGs) all at $z=2.67$ (within $\Delta V=800$ km s$^{-1}$) spanning a projected extent of $>300$ kpc lensed by a $z=0.77$ foreground cluster. NIRCam and ALMA Band 6 continuum measurements map the stellar distribution and thermal dust emission respectively at a spatial resolution of $\sim$0.15$^{\prime\prime}$. This analysis reveals a diverse population of DSFGs, with evidence of both interacting and non-interacting systems exhibiting a wide range of morphological features including spiral arms, bars, bulges, clumps/stellar clusters, tidal tails/debris and  displaced molecular gas reservoirs. Comparing the rest-frame J- band continuum (F444W) vs (i-J) color  (F277W$-$F444W), we find a wide range of values, suggesting a $>$1-dex spread in stellar mass and a dust attenuation reddening of $\Delta$A$_\textup{v}>1$ mag. The DSFG members exhibit varying dust sizes relative to the stellar emission, ranging from compact dusty cores to galaxy-wide emission. Resolved color maps of individual sources showing a spread as high as F277W$-$F444W$=2$ mag suggesting complex stellar-to-dust geometry. Although gas-rich mergers are identified in the core, the most red and dust emitting members are disks exhibiting clumpy structure indicating secular growth can drive these starburst events. Such a remarkable range in properties within this sample suggest DSFGs in protocluster core environments follow diverse evolutionary pathways towards their transition into quiescent, elliptical cluster galaxies. 



\end{abstract}


\section{Introduction} \label{sec:intro}


Protoclusters are the progenitors of galaxy clusters, the most massive gravitationally bound structures in the Universe, and host the stellar nurseries that form the most massive galaxies within their central cores \citep{Overzier_2016, Chiang_2013}. At $z>2$, protocluster cores have been identified to be very active sub-mm bright sources; driven by overdensities of dusty star-forming galaxies (DSFGs) \citep{Casey_2016, alberts_2022}. These structures typically harbor tens of DSFGs within compact regions spanning a few hundred projected kpc, with total star formation rates (SFR) of thousands of solar masses per year \citep[e.g.,][]{Wang_2016, Miller_2018, Oteo_2018, Foo_2025}. The rapid stellar build-up in these extreme systems can drive the formation of brightest cluster galaxies (BCGs) in cluster cores by $z\lesssim1$ \citep{Webb_2015, Rennehan_2020}. Yet, the underlying physical mechanisms that trigger and regulate the stellar growth of DSFGs in protocluster cores are unsettled. 
\par
Based on Hubble Space Telescope (HST) studies, merger events have been proposed to be a driving mechanism of DSFGs \citep{Chen_2015}. However, such morphological rest-frame UV/optical classifications were susceptible to dust attenuation and were unable to fully probe the distribution of stellar versus thermal dust emission from obscured star formation \citep{Simpson_2017}.  James Webb Space Telescope (JWST) observations now enable the characterization of dusty stellar continua at $z>2$ to at high spatial resolution. Several JWST DSFG studies have since measured merger fractions comparable to their general galaxy population \citep{McKay_2025, Ren_2025}, while others have found that the majority of their sample shows clear evidence of interactions \citep[e.g.,][]{Hodge_2025}. Nonetheless, such classifications consist of limited sample sizes and often rely on visual identification. 
\par
Interferometric kinematic observations of sub-mm interstellar medium (ISM) gas line tracers (e.g., \textsc{[C\,ii]}, CO) have revealed rotationally dominated systems with low levels of turbulence in DSFG disk samples \citep[e.g.,][]{Rizzo_2020,Rizzo_2021,Roman-Oliveira_2023,Rizzo_2023, Amvrosiadis_2025}, demonstrating such star formation can be sustained via secular processes without mergers. Recent JWST studies have identified different disk-like stellar morphology in DSFGs including evidence of spiral arms, galactic bars and bulge formation \citep[e.g.,][]{Chen_2022,  Hodge_2025, Umehata_2026}. The wide-ranging observations on DSFG properties suggests that they are comprised of a heterogeneous population representing different evolutionary pathways and origins \citep{Hayward_2011, Cooper_2025}. Given DSFGs are linked to large-scale structure assembly \citep{Casey_2016, Chapman_2009}, connecting how protocluster environments influence their properties is essential for tracing their evolution.
\par
Multi-wavelength observations on DSFG-rich protocluster cores have demonstrated that these environments can accelerate galaxy evolution, finding evidence of `top-heavy' stellar initial mass functions and signatures of early transition toward massive, quiescent galaxies \citep[e.g.,][]{Sun_2024, Sillassen_2026, Witten_2026}. Given the overdense nature of protocluster cores, galaxy mergers have been shown to be key drivers for the stellar build-up in the formation of large-scale structure \citep{Hine_2016, S.Liu_2023}. Yet, the efficient accretion of gas onto large-scale filaments \citep{Dekel_2009, Daddi_2021} and/or the recycling of gas within the circum-galatic medium (CGM) \citep{Emonts_2018} may sustain more secular star formation in disk galaxies. Recent JWST/ALMA observation have shown evidence that disk-like morphologies (with central bulge growth) \citep{Umehata_2025,Zhang_2026, Umehata_2026} and kinematic properties \citep{Venkateshwaran_2024} may dominate the DSFG population in protocluster environments. Given the limited sample size and complexities in isolating different environmental effects, the evolution of DSFG in protocluster cores remains difficult to constrain. 

\begin{figure*}
	\centering\includegraphics[scale =0.61]{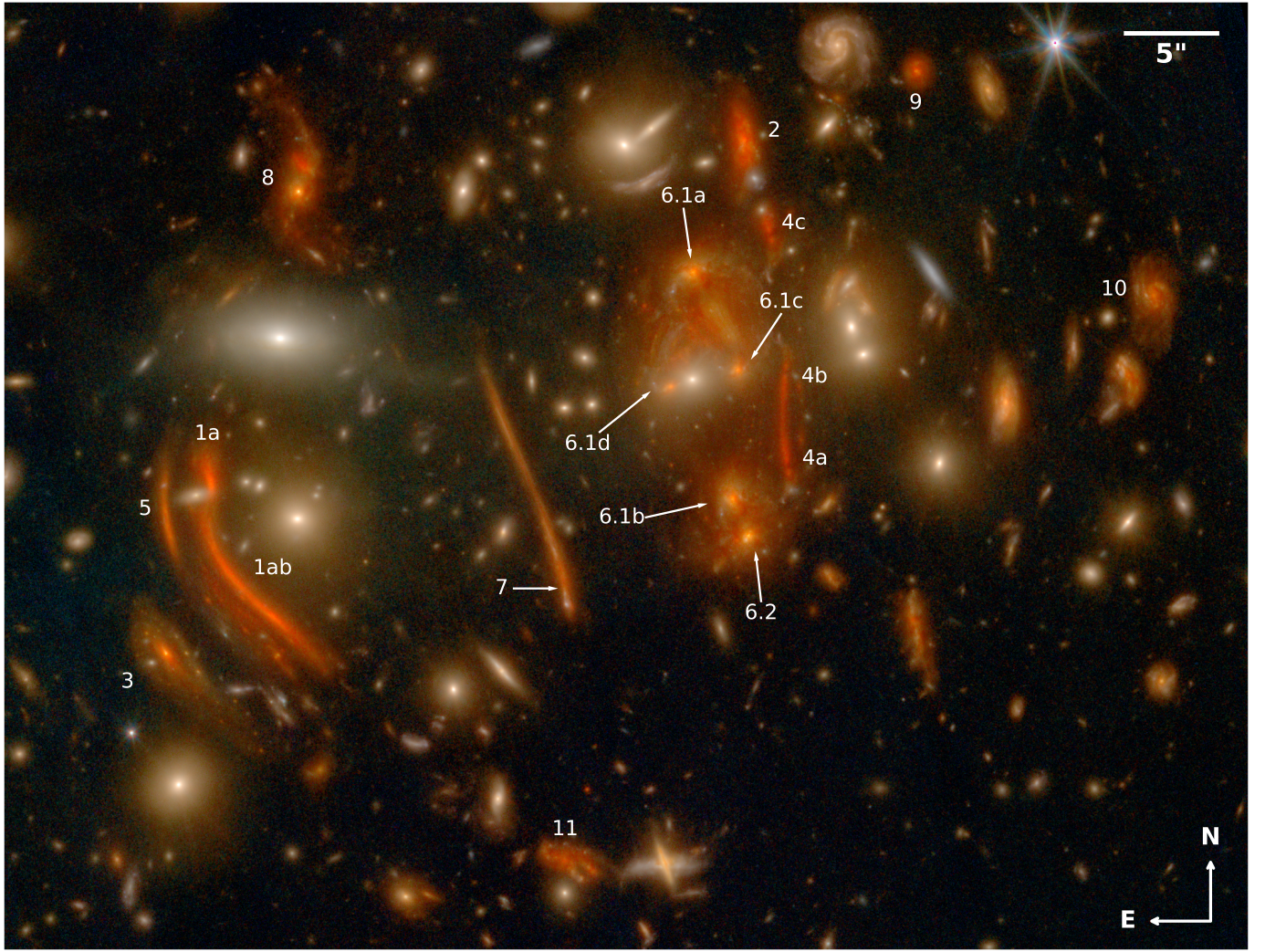}  
	\caption{The RGB six color composite color image produced from NIRCam filters (Blue: F090W, F115W, Green: F150W, F277W, Red: F356W, F444W) of the J0846 lensed protocluster core field. NIRCam resolves the dusty stellar distribution of 11 DSFGs at $z=2.67$, being strongly lensed by a foreground cluster. }
\label{fig:1}
\end{figure*}

\par
The PJ0846 +15 (J0846, RA: 08:46:49, DEC: 15:05:58) field was first identified within a larger parent sample in the \textit{Planck} All-Sky Survey to Analyze Gravitationally-lensed Extreme Starbursts \citep[PASSAGES][]{Harrington2016, Berman2022}. From the initial $2.5'$ \textit{Planck} detection, follow-up AzTEC 1.1mm and single-dish multi-J CO line observations resolved two distinct sources \citep{Harrington2021, Berman2022}. 
Subsequent ALMA observations revealed that J0846 did not consist of just a single sub-mm source, but instead of at least $11$ DSFGs at $z=2.67$ detected via CO(3–2) line detections \citep{Foo_2025}. Various optical/near-IR imaging and spectroscopy of the field enabled the construction of a lens model for the foreground cluster at $z=0.766\pm0.002$. Based on this model, these 11 DSFGs span a projected physical extent of $\lesssim300$ kpc with a total infrared-based $\textup{ SFR}=5200^{+3200}_{-2000} \textup{ M}_\odot$ yr$^{-1}$ \citep{Foo_2025} corrected for magnification. J0846 is thus one of the most active sites of star formation discovered at $z>2$, while its alignment with a foreground cluster lens provides a magnified glimpse into the rapid stellar build-up within a nascent cluster core. 
\par
In this paper we present new JWST NIRCam and ALMA continuum observations to study the stellar structure and dust continuum of this DSFG-rich lensed protocluster core. This manuscript is organized as follows. In Section \ref{Sec:Observations} we describe the JWST/NIRCam and ALMA band 6 continuum observations. In Section \ref{Section:Results} we present the stellar near-infrared (NIR) and dust continuum properties of the J0846 DSFG members. In Section \ref{Section:Discussion} we then further discuss their resolved morphological features and implications on the stellar build-up within protocluster core environments. We summarize our findings and discuss the outlook in Section \ref{Sec:Summary}. This paper assumes a flat $\Lambda$CDM cosmology with $H_0=70$ km s$^{-1}$ Mpc$^{-1}$, $\Omega_{m}=0.3$, $\Omega_{\Lambda}=0.7$ and 1$^{\prime\prime}$=8.107 kpc at $z=2.67$. JWST NIRCam imaging unveils the spectacular lensing arcs of these dusty galaxies, earning its name as the \textit{Koi Pond} protocluster.


\section{Observations} \label{Sec:Observations}

\subsection{JWST/NIRCam}

In this work, we analyze JWST/NIRCam observations of the J0846 field obtained through the GO Cycle 3 program PID 6782 (PI: Foo). Observations were carried out on UT 2025 March 19 in six filters reaching total exposure times of 2920s for the F090W/F444W pair, 1890s for the F115W/F277W pair, and 1460s for the F150W/F356W pair. The 5$\sigma$ point-source depths reached for the six filters, from shortest to longest wavelengths, are 28.62, 28.48, 28.55, 28.48, 28.54, 28.71 mag. 
We reduced the NIRCam data using version 1.16.1 of the JWST Science Calibration Pipeline (CALWEBB; STScI), with calibration reference files selected via the Calibration Reference Data System (CRDS) context \texttt{jwst\_1303.pmap}. 
To mitigate \texttt{JUMP} artifacts 
 during stage 1, we customized the standard pipeline and its three stages following the PEARLS 
\citep{Windhorst_2023} prescriptions. We removed both $1/f$ noise and ``wisps'' (i.e. straylight artifacts on the A1, A2, B1, and B2 shortwavelength detectors) by applying the JumProPe \citep{DSilva2025, Robotham2023} suite on the calibrated exposure files from stage 2. Astrometry of the image mosaic was corrected to the WCS alignment of GAIA DR2 \citep{Gaia_2023}. 
We also define the \texttt{resample} step to produce mosaics drizzled on 0\farcs{03}/pix grid across the six filters all in the north-up direction. 
\par
To conduct PSF-matching we utilize \texttt{WebbPSF} \citep{Perrin_2014} to generate PSF models and the \texttt{PyPHER} \citep{Boucaud_2016} tool to make the convolution kernels matching all filters to the F444W image. To measure photometry on the PSF-matched NIRCam images we produced custom, manually drawn apertures based on the F444W filter. This approach was utilized to account for the crowded nature of the field as well as the complex morphologies of merger systems and due to strong lensing. In this initial analysis we do not subtract any intracluster foreground light, although it is evident that significant contamination may be present for the ID6.1/6.2ab images. Therefore, we only report the photometry of the ID6.1/6.2c images for that source.





\begin{figure*}
	\centering\includegraphics[scale =0.37]{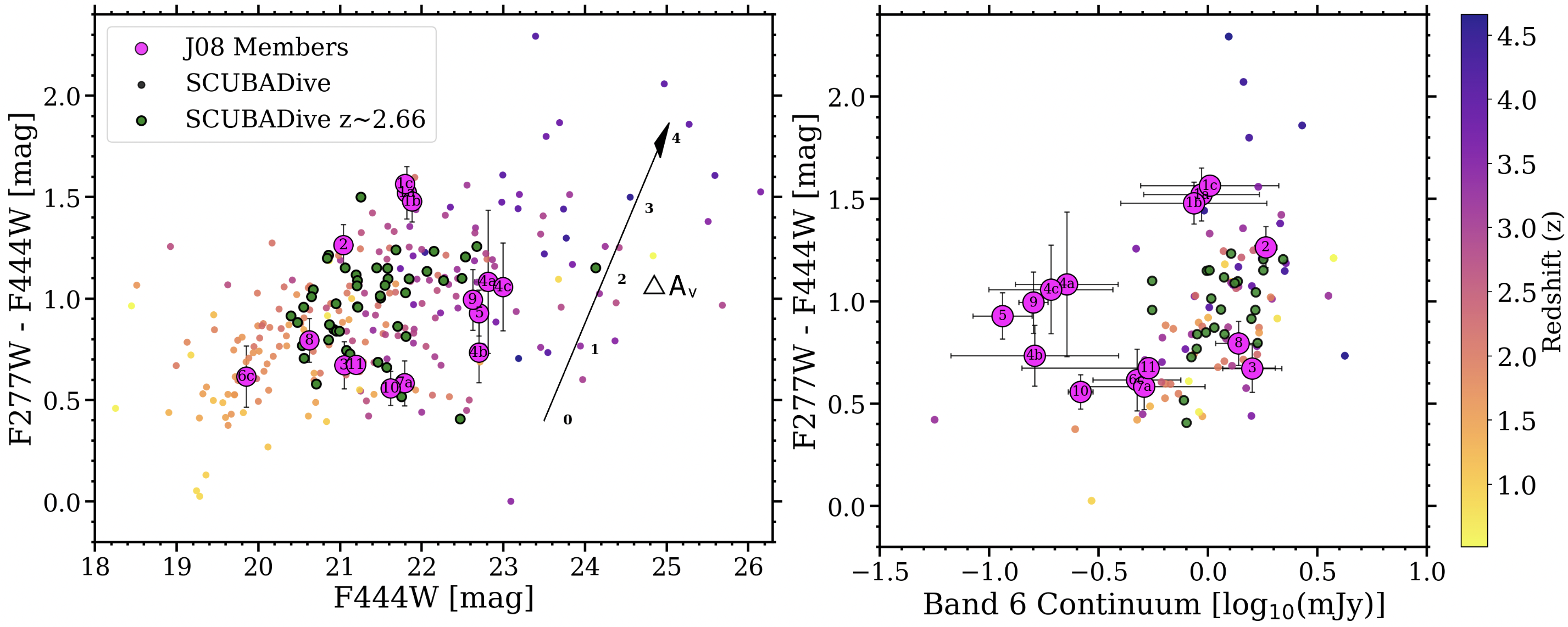} 
	\caption{ \textit{Left Panel:} Color-magnitude diagram utilizing F277W and F444W filters showing the relationship between the stellar color and continuum. We show the expected dust reddening vector ($\Delta$A$_V$) derived from the Calzetti attenuation curve for a redshift of $z=2.67$. \textit{Right Panel:} F277W and F444W filters versus ALMA band 6 dust continuum flux densities. All measurements of the J0846 lensed members are magnification corrected based on the lens model presented in \cite{Foo_2025}. For comparison we plot the sub-mm selected SCUBADive sample, where the green points are sources at redshift range of $z=2.67 \pm 0.2$ \citep{McKinney}. }
	\label{fig:2}
\end{figure*}

\subsection{ALMA}
ALMA continuum observations of the J0846 field were obtained through a Cycle 10 program (2023.1.00299.S. PI: Foo) to follow-up Cycle 5 ALMA observations presented in \cite{Foo_2025}. We designed the Band 6 observations to map the dust continuum of all confirmed protocluster members, using a 27-pointing mosaic to cover a $60^{\prime\prime}\times60^{\prime\prime}$ area encompassing the critical curve of the foreground cluster lens. The frequency setup was centered at 228 GHz (1315 $\mu$m) corresponding to rest frame 358 $\mu$m at the redshift of the protocluster core ($z=2.67$). Observations were obtained in the 12 m array in configuration 3 and 6 (i.e., TM2 and TM1), with an average on-source integration time of 330 (C3) and 881 (C6) seconds, respectively. We reduced the dataset with the Common Astronomy Software Application \citep[CASA,][]{McMullin2007, CASA_2022} version 6.5.4.9 with the ALMA pipeline heuristics \citep{Hunter_2023}. Using the \texttt{tclean} routine in the CASA reduction pipeline we produce multi-frequency synthesis continuum imaging including only the line-free channels to avoid contamination to the dust emission. To account for the compact and diffuse emission of these lensed sources we utilize the multi-scale deconvolver algorithm. We set scale sizes to be $[0, 5, 15, 20, 25, 30, 35, 40, 45, 50, 60, 75]$ pixels, cleaning down to a threshold of 1.5$\sigma$ using a Briggs weighting scheme with a robust parameter of 0.5. Masks for cleaning are generated via the \texttt{auto-multithresh} algorithm \citep{Kepley_2020}. For this work, TM1 and TM2 observations are imaged independently. Flux densities are reported from the TM1 produced continuum images given the observations are deeper than taken for TM2. A comparison of the flux densities measured in TM1 and TM2 utilizing equivalent custom apertures shows agreement to within 10\%, consistent with the nominal flux uncertainties of ALMA \citep{Francis_2020}, however imaging combination with TM1 and TM2 mosaics will be the focus of future work.
\par
For TM1, the image pixel size is $0.03^{\prime\prime}$ with a restoring beam size of $0.17^{\prime\prime}\times0.17^{\prime\prime}$ and position angle of $-48 ^\circ$. The final root mean square (rms) sensitivity achieved for the continuum imaging was $\sim0.034$ mJy beam$^{-1}$. For TM2, the image pixel size is $0.12^{\prime\prime}$ with a restoring beam size of $0.56^{\prime\prime}\times0.52^{\prime\prime}$ and position angle of $-32 ^\circ$. The final rms sensitivity achieved for the continuum imaging was $\sim0.055$ mJy beam$^{-1}$. To measure continuum flux densities, the same custom apertures were utilized for the NIRCam photometry based on the F444W filter image. For all sources the stellar continuum F444W based custom apertures completely encapsulated the corresponding dust emission down to at least $2\sigma$ (i.e., stellar sizes were always larger than dust continuum sizes). 



\section{Results} \label{Section:Results}

\begin{deluxetable*}{cccccccccc}
\tablecaption{Intrinsic Properties of the $z=2.67$ DSFG Members}
\label{Table:1}
\tablecolumns{10}
\tablewidth{0pc}

\tablehead{
  \colhead{ID}  & 
  \colhead{x}   & 
  \colhead{y}   & 
  \colhead{$z$} & 
  \colhead{F444W} & 
  \colhead{F277W$-$F444W} & 
  \colhead{$\mu$} & 
  \colhead{$S_{\text{dust}}$} & 
  \colhead{Int/Merg} & 
  \colhead{Notes} \\
  \colhead{} & 
  \colhead{arcsec} & 
  \colhead{arcsec} & 
  \colhead{} & 
  \colhead{[AB mag]} & 
  \colhead{[AB mag]} & 
  \colhead{} & 
  \colhead{[mJy]} & 
  \colhead{} & 
  \colhead{}
}

\startdata
1$^a$  & 3.7   & $-4.8$ & 2.663 & $21.79 \pm 0.02$ & $1.57 \pm 0.04$ & $8.2 \pm 5.0$  & $0.93 \pm 0.57$ &  &  clumpy  \\ 
2      & 18.9  & 3.7    & 2.665 & $21.04 \pm 0.04$ & $1.26 \pm 0.10$ & $4.2 \pm 0.5$  & $1.83 \pm 0.22$ &  & clumpy \\ 
3      & $-0.2$ & $-7.5$ & 2.660 & $21.05 \pm 0.06$ & $0.67 \pm 0.12$ & $3.2 \pm 1.0$  & $1.59 \pm 0.50$ & $\checkmark$ & tidal tail, clumps, bulge \\ 
4$^a$  & 19.6  & 1.5    & 2.669 & $22.99 \pm 0.11$ & $1.06 \pm 0.22$ & $3.7 \pm 2.0$  & $0.23 \pm 0.12$ & $\checkmark$ & CO(3–2) gas rotation \\ 
5      & 1.6   & $-5.6$ & 2.666 & $22.70 \pm 0.06$ & $0.93 \pm 0.11$ & $6.4 \pm 2.0$  & $0.11 \pm 0.04$ &  & smooth disk \\ 
6$^a$  & 17.5  & $-0.1$ & 2.666 & $19.85 \pm 0.09$ & $0.62 \pm 0.15$ & $2.6 \pm 1.2$  & $0.47 \pm 0.22$ & $\checkmark$ & tidal debris, bulges, gas offset \\ 
7      & 12.8  & $-1.5$ & 2.666 & $21.79 \pm 0.06$ & $0.58 \pm 0.11$ & $3.1 \pm 2.0$  & $0.51 \pm 0.33$ &  & UV-bright clump\\ 
8      & 7.2   & 0.8    & 2.662 & $20.62 \pm 0.06$ & $0.80 \pm 0.11$ & $2.9 \pm 0.7$  & $1.38 \pm 0.33$ & $\checkmark$ & tidal tails, PSF-signature \\ 
9      & 26.2  & 6.5    & 2.669 & $22.62 \pm 0.08$ & $0.99 \pm 0.15$ & $2.6 \pm 0.4$  & $0.16 \pm 0.02$ &  & bulge, spiral arms \\ 
10     & 34.0  & 2.7    & 2.668 & $21.62 \pm 0.05$ & $0.56 \pm 0.08$ & $2.3 \pm 0.3$  & $0.26 \pm 0.03$ &  & bulge, spiral arms, bar \\ 
11     & 12.7  & $-12$  & 2.664 & $21.20 \pm 0.02$ & $0.67 \pm 0.04$ & $1.5 \pm 2.0$  & $0.53 \pm 0.71$ &  & bulge, spiral arms, clumps \\ 
\enddata

\tablecomments{Column 2, 3: Angular separation from the center$^b$, where +x and +y are in the North and West direction, respectively; Column 5: De-lensed F444W magnitude; Column 6: De-lensed F277W$-$F444W color; Column 7: Magnification factor ($\mu$); Column 8: De-lensed intrinsic ALMA Band 6 dust continuum flux ($S_{\text{dust}}$); Column 9: Evidence of interaction/merger}
\tablenotetext{a}{NIRCam photometry and positions are measured from sub-image c (1c, 4c, 6c) as detailed in text, while corresponding dust fluxes follow specific component availability mapping (1c, 4c, 6c respectively) matching the underlying magnification constraints.}
\tablenotetext{b}{Center position: RA=131.70877, DEC=15.09936.}
\end{deluxetable*}




In Fig.~\ref{fig:1}, we present the 6-band NIRCam color composite image of the J0846 lensed protocluster core, revealing a \textit{Koi Pond} of resolved dusty stellar emission. These NIRCam observations probe the rest-frame 0.245–1.21 $\mu$m emission, covering longer wavelengths that are less affected by dust obscuration than the previous HST F160W imaging. ALMA band 6 observations map the rest frame 358 $\mu$m thermal dust continuum at similar $\sim0.15^{\prime\prime}$ angular resolution. We now are able to resolve the stellar and dust structure of all 11 galaxies previously confirmed in CO gas emission \citep{Foo_2025}. 


\subsection{NIR Color-Magnitude Properties} \label{Stellar Reddening}

We construct F277W and F444W color-magnitude diagram as a proxy to infer the relative rest-frame NIR stellar reddening among the DSFG members (Left panel in Fig.~\ref{fig:2}). We select F277W$-$F444W as it is redward of the Balmer break, thereby avoiding the color degeneracy. We measure a sample median stellar continuum of F444W$=21.62^{+1.03}_{-0.75}$ mag, ranging F444W$=20-23$ mag and color of F277W$-$F444W$=0.80^{+0.34}_{-0.20}$ mag, ranging F277W$-$F444W$=0.5-1.6$ mag. All flux measurements have been corrected for lensing magnification based on the model presented in \cite{Foo_2025} (assuming lensing magnification is equal across both filters), and should improve based on the construction of the refined JWST-based lens model (Foo et al. in prep.). Measurements for each member are presented in Table~\ref{Table:1}. 

\par
We compare this to the SCUBADive sample of 289 DSFGs in the COSMOS field presented in \cite{McKinney}, with similar NIRCam and ALMA band 6/7 observations (Fig.~\ref{fig:2}), and redshift of $\left\langle z \right\rangle=2.6^{+1.0}_{-0.8}$ comparable to J0846. Within the same redshift of $z=2.67\pm0.2$, the members share a similar color-magnitude space with the SCUBADive subsample, where they measure $\left\langle A_{V} \right\rangle=2.6^{+1.6}_{-1.1}$ and $\langle \log(M_{*} / M_{\odot}) \rangle = 11.2^{+0.4}_{-0.5}$. We also plot the expected ${A}_{V}$ dust reddening vector derived from the Calzetti attenuation curve \citep{Calzetti_2000} for the redshift of $z=2.67$, assuming the same stellar populations. These results suggest the J0846 members are massive ($M_{*}>10^{10} M_{\odot}$) DSFGs with a spread in stellar masses of at least $>1$ dex (see Figure 3 in \citealt{Orozco_2026}) and dust attenuation of ${A}_{V}>1$ mag. Future work will conduct stellar spectral energy distribution analysis, as without mid-infrared photometry stellar mass estimates of high$-z$ red galaxies could be overestimated up to 1 dex \citep{Williams_2024}. 
\par
In Figure \ref{fig:3}, we provide a detailed view of the multi-band stellar and cold ISM emission (dust and CO gas kinematics) of the individual DSFGs in J0846. We show the corresponding color maps by taking pixel-by-pixel F277W$-$F444W measurements of the PSF-matched images, showcasing the variations in color across each member. In the sample we find a mix of shallow to steep color gradients on resolved scales, demonstrating  aggregate measurements may not accurately reflect the overall reddening properties. Systems with complex and extended morphologies (i.e., merger systems, ID8 and ID6, see Section \ref{Section-Mergers}), exhibit central stellar components that are redder up to F277W$-$F444W$\sim2$ mag than the diffuse and faint surface brightness emission detected in tidal features. We also find evidence of patchy dust attenuation as we detect rest-frame UV/optical emission (ID4, ID6, ID7, ID11) that can penetrate the ISM.




\begin{figure*}
	\centering\includegraphics[scale =0.45]{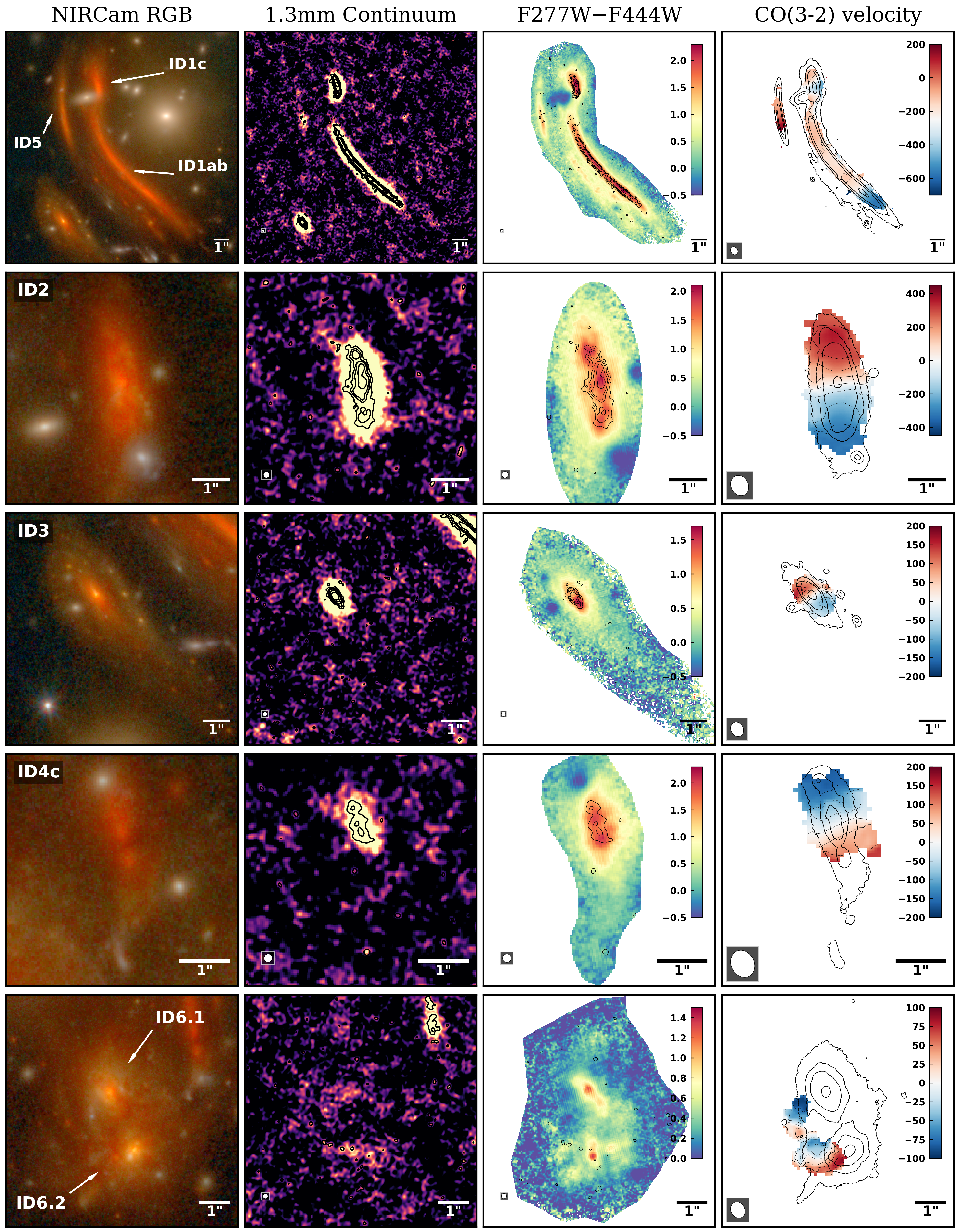}
	\caption{Individual image stamps of the J0846 members. In the 1st column to the left shows the composite NIRCam color image. The 2nd column shows band 6 dust continuum feathered image, with the TM1 contours overlaid (and restoring beam size). In the 3rd column we display the F277W$-$F444W color maps with TM1 dust continuum contours overlaid. In the 4th column are the CO(3–2) velocity maps (in km s$^{-1}$) from \citep{Foo_2025} with the F444W stellar continuum contours overlaid in black. NIRCam and ALMA observations highlight the different morphologies, and stellar color relative to their thermal dust emission at sub-kpc scales. 
    }
	\label{fig:3}
\end{figure*}

\begin{figure*}
    \figurenum{\ref{fig:3}} 
	\centering\includegraphics[scale =0.45]{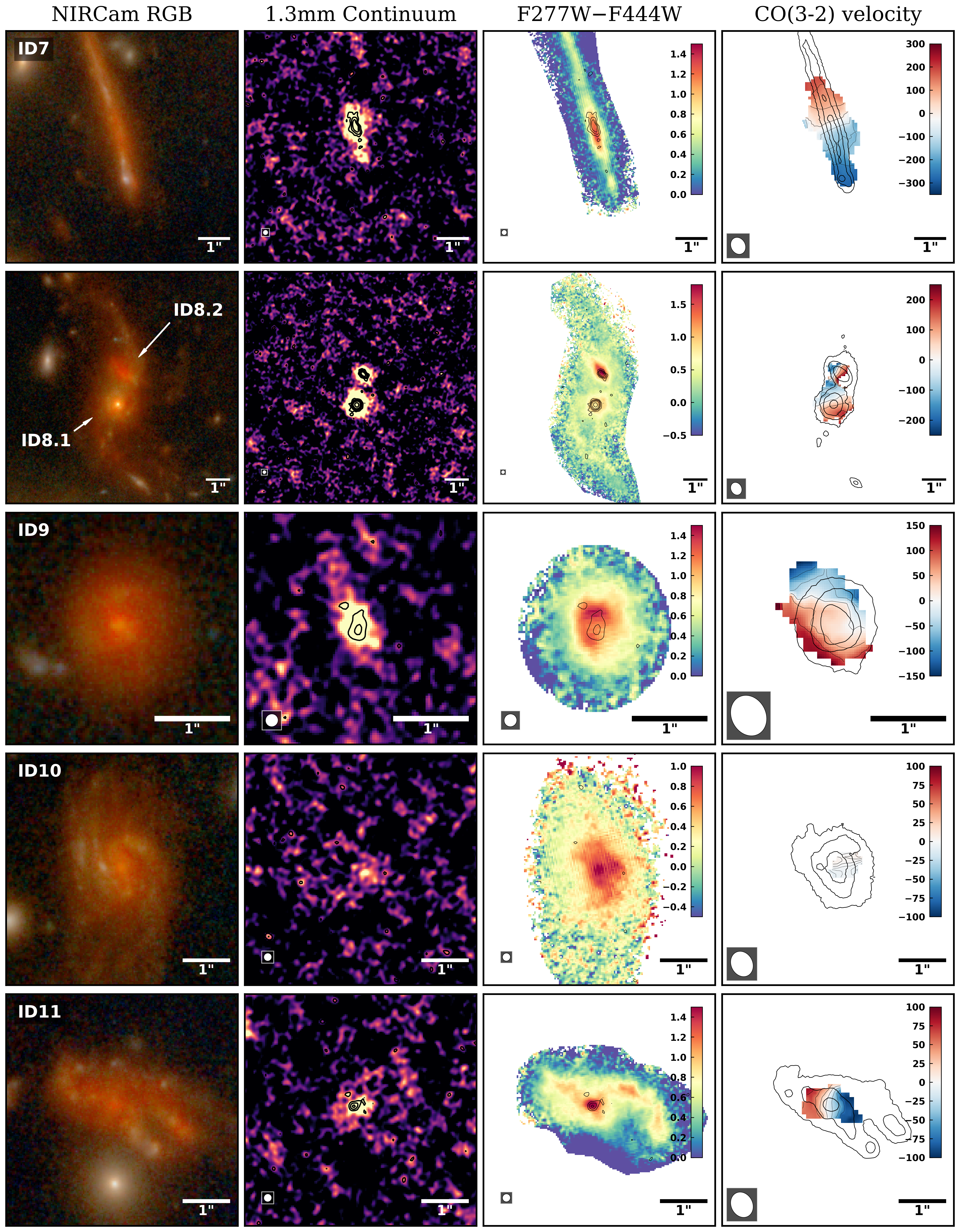}
	\caption{Continued
    }
\end{figure*}

\subsection{Comparing ALMA and NIRCam Continuum} \label{Stellar Reddening}

\begin{figure*}
	\centering\includegraphics[scale =0.39]{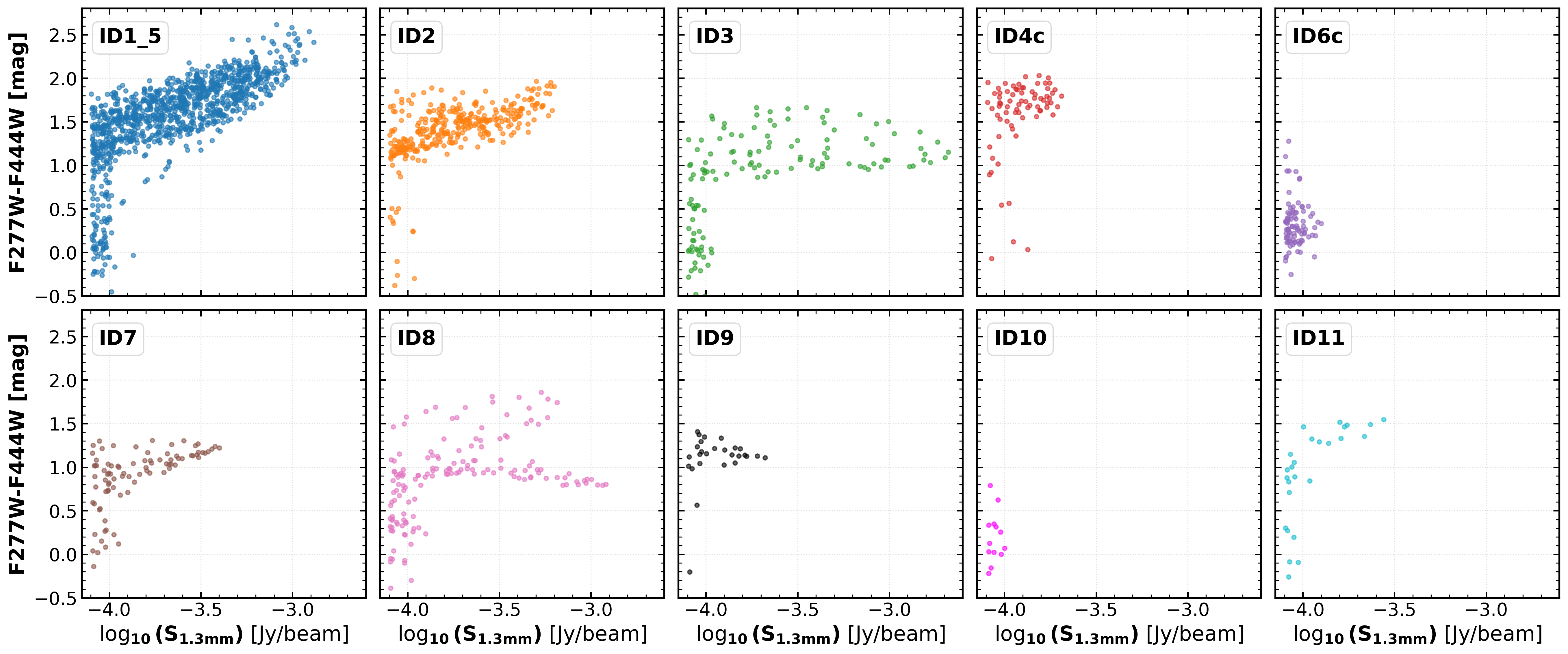} 
	\caption{ Resolved F277W$-$F444W color excess versus the band 6 dust continuum surface brightness for each member. A SNR$>2\sigma$ cut in 1.3 mm flux density (measured via $2\times2$ pixel binning) is applied, where NIRCam generally probes the more extended stellar continuum where dust emission is not detected. Correlation between F277W$-$F444W and dust continuum suggest their color excess is predominantly driven by dust reddening rather than stellar age. However, the diversity in profile shapes indicate different dust-to-stellar distribution among the members.  }
	\label{fig:4}
\end{figure*}

The J0846 DSFGs exhibit a wide range of flux densities in the ALMA 1mm mosaic, of $S_{1.3\textup{mm}}\simeq0.1 - 1.9$ mJy after correcting for lensing magnification. In the right panel of Fig.~\ref{fig:2} we show F277W$-$F444W versus $1.3$mm flux densities, comparing to the 105/289 sources with Band 6 observations in the SCUBADive sample. A handful of relatively faint sources are uncovered with $<0.3$ mJy, where lensing enables the detection of this low-dust-continuum DSFG population within J0846. Although aggregate measurements do not exhibit any strong correlation between dust continuum and F277W$-$F444W in Fig.~\ref{fig:2}, the resolved maps in the 3rd column of Figure \ref{fig:3} generally show that the F277W$-$F444W color excess is closely associated with the observed dust emission. This may suggest that the red color excess is driven by dust reddening rather than stellar age \citep{Hodge_2025}, however this does not account for any degeneracies in star-formation history and metallicity. In Figure \ref{fig:4} we plot the resolved F277W$-$F444W measurements versus dust continuum surface brightness within each member, finding a similar trend for several sources (ID1, ID2, ID4, ID5, ID7). However, there are exceptions observed in ID8 (specifically for ID8.1) and ID3 (exhibiting the brightest peak of $\sim2.3\times10^{-4}$ Jy beam$^{-1}$). For both sources the 1.3mm continuum peaks are aligned with their optically bright point-like cores, likely driven by central starbursts or AGN able to emerge from its dust envelope (see Section \ref{Resolving the Stellar Build-up of DSFG Members}). Nonetheless, the scatter in data points for each source varies, suggesting different dust-to-stellar distribution among this population of DSFGs.

\par

To better visualize the dust continuum we produced a combined TM1 and TM2 feathered image using the CASA task \texttt{feather} \citep[i.e.,][]{Cotton_2017} with a an arbitrary flux scaling \texttt{sdfactor=3}, providing a detailed view of the multi-scale emission morphology. We present the combined feathered imaging for each DSFG member in Figure \ref{fig:3} with TM1 contours overlaid for comparison. 
We first visually examine the image-plane properties, finding overall that the relative sizes between the F444W stellar continuum and 1.3 mm dust components show significant diversity in the sample. Some members host compact dust components residing exclusively within their central cores (ID10, ID11), while others exhibit more wide-spread dust continuum emission (ID1, ID2). Though, in all source the dust continuum remains more compact to their stellar counterparts, consistent with other JWST-ALMA studies \citep[e.g.,][]{Chen_2022, Gillman_2024, Hodge_2025}. We typically find that the dust and stellar peaks are co-located except in ID6, which exhibits a pair of merging galaxies (see Section \ref{Resolving the Stellar Build-up of DSFG Members}). The agreement in flux densities between TM1 and TM2 would indicate that minimal emission is resolved out from large-scale structure. This may suggest dust structure is more compact and clumpier than expected \citep[e.g.,][]{Hodge_2019}, where observations with insufficient resolution may overestimate continuum size measurements. TM1 imaging of J0846 exhibit smaller dust continuum sizes and is able to resolve individual $\lesssim$ kpc-size clumps in several members (ID2, ID4, see Section \ref{Resolving the Stellar Build-up of DSFG Members}). 

\section{Discussion} \label{Section:Discussion}

CO observations in \cite{Foo_2025} provided kinematic evidence for rotating molecular gas disks, as well as mergers and interactions among, the DSFG members. The majority of the members lie within $<20$ kpc and $\Delta V <500$ km s$^{-1}$ of another galaxy, and are therefore expected to undergo some merger event if not already interacting \citep[e.g.,][]{Man_2016}. We highlight that molecular gas rotation is not exclusive to secular evolution \citep[e.g.,][]{Ueda_2014}, as this protocluster environment  hosts disks with and without signs of interaction. This section we  discuss the NIRCam and Band 6 ALMA observations that now resolve a wide range of morphological features as well as its implications on DSFGs evolution within protocluster core environments.

\subsection{Evidence for Mergers and Interactions } \label{Section-Mergers}

We find evidence of different merger/interaction scenarios and stages among four different members (ID3, ID4, ID6, ID8, see Figure \ref{fig:3}). ID8 consists of two kinematically distinct CO(3–2) components (8.1 and 8.2, with a $\sim100$ km s$^{-1}$ offset), also discernible in the Band 6 dust continuum emission. The less CO-luminous component, 8.2, exhibits a CO(3–2) line-profile with a large FWHM of $\sim800$ km s$^{-1}$ suggesting highly turbulent/disturbed gas and/or an outflow. The NIRCam imaging now detects the two merging stellar counterparts exhibiting different UV-to-optical colors. 8.2 exhibits a more smooth optically-dark, redder component, whereas 8.1 hosts a UV-bright central point source with what appears to be a PSF signature indicating a bright nuclear starburst or potential AGN/quasar hosts triggered by the merger event. Most prominently it reveals extended tidal “antennae” tails  \citep[e.g.,][]{Toomre_1972, Duc_2013, Wen_2016} of $\sim8$\,kpc in length, whereas the CO(3—2) and dust continuum emission is not as extended and remains more compact. ID8 may be an early–to–intermediate stage prograde merger, likely after the first pericenter passage but before final coalescence given the two distinct components.


ID3 exhibits diffuse extended emission resembling a tidal tail trailing in the south direction, though this feature may be due to lensing. The color difference maps (Fig. \ref{fig:3}) reveal a red component offset from the brightest peak of the stellar and dust/CO emission. This could be a stellar companion undergoing final coalescence during a late-stage merger. We detect $>24$ prominent point sources scattered throughout the extended diffuse stellar emission (though at least one of which is likely a foreground galaxy), some of which reside in the outskirts of the potential tidal tail. These objects may represent stellar clusters that have been tidally ejected or accreted during the merger event (e.g., \citealt{Claeyssens_2023, Whitaker_2026}, see Section \ref{Resolving the Stellar Build-up of DSFG Members}).

The stellar morphology of ID4 exhibits two merging components and significant color variation, with the rest-frame UV/optical emission spatially offset from the dust/gas (see Figure \ref{fig:3}). Despite the apparent merger components, the CO(3–2) velocity map reveals an ordered velocity gradient suggestive of a rotating gas disk, while the 1.3 mm dust-continuum similarly shows no evidence of multiple components (though it does exhibit clumpy substructure). However, given insufficient resolution and/or particular sight lines, kinematic observations may not be able to distinguish between the relative motion of merging galaxies and the rotation of a single disk \citep[e.g.,][]{Simons_2019, Fraternali_2021}. Alternatively, the rotating molecular gas disk may have survived the merger event \citep[e.g.,][]{Ueda_2014}.

The CO(3–2) emission of the ID6 merger system similarly consist of two distinct components (6.1 and 6.2), with a velocity offset of $\sim150$ km s$^{-1}$. NIRCam imaging reveals the merging pair of stellar components (see Fig. \ref{fig:3}). The quadruply imaged Einstein cross, initially identified as ID12 in \cite{Foo_2025} is confirmed to be the stellar counterpart of 6.1. Its merging companion, 6.2, is approximately $\sim 5$ kpc away (corrected for magnification), and does not fully reside inside the lensing caustic and thus does not produce multiple images (see Fig. \ref{fig:A1} in the Appendix). Both galaxies exhibit spheroidal stellar structures with bulge-like central features. In contrast to the ID8 merger, it lacks defined extended tails and instead exhibits more diffuse emission with potential tidal shells. Barring any projection effect, it may indicate ID6 is a more radially-dominated collision, and/or consist of more kinematically slowly rotating galaxies \citep{Quinn_1984, Hernquist_1992, Valenzuela_2024}. 

\subsubsection{Tidally Displaced Gas Reservoir?}

Strikingly, for both galaxies comprising ID6, the gas and dust emission is offset $\sim0.8^{\prime\prime}$ ($\sim2.5$ kpc, magnification corrected) from the stellar emission, instead coinciding with the tidal debris (see far right panel in Figure \ref{fig:3}). Much of this emission originates from the ID6COab image that resides on the lensing caustic between the two merging galaxies providing a highly magnified view ($\mu\sim80$) of this intermediate tidal feature (see Fig. \ref{fig:A1} in the Appendix). Although the displacement of this molecular gas is likely driven by tidal interactions, it may not necessarily represent newly formed tidal gas reservoirs (i.e., self-gravitating). The CO(3–2) velocity field suggests 6.1 and 6.2 are two dynamically distinct structures (with a large PA difference), both exhibiting rotation resembling gas disks. Instead, this may indicate that sufficient angular momentum from their progenitor systems was retained, and that their disks were not completely destroyed by the tidal forces of the interaction \citep[e.g.,][]{Hopkins_2009}. However, it is unclear whether this gas reservoir can be displaced by such a magnitude (several kpc) while still exhibiting evidence of coherent rotation. Ram-pressure stripping may also be at play, given its residence in this protocluster core environment \citep[e.g.,][]{Noble2019,Cramer_2023,xu_2026, Zhou_2026}. The mechanisms driving this lensed merger event will be further explored in future work.

\begin{figure*}
	\centering\includegraphics[scale =0.34]{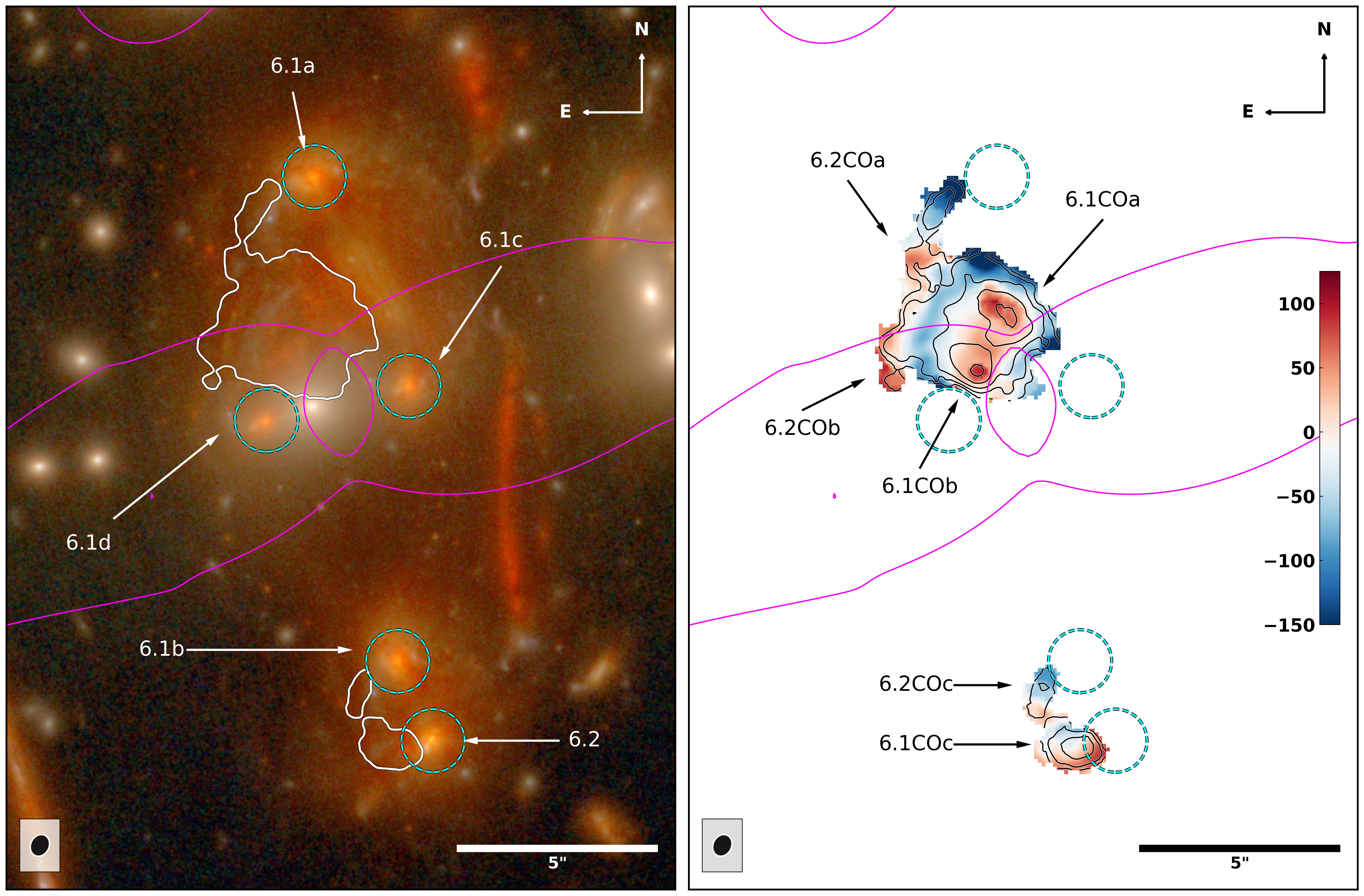}  
	\caption{Zoom-in of the quadruply-imaged ID6 merger system. On the left panel shows the NIRCam color image with the CO(3–2) moment-0 contour overlaid in white. The right panel shows the CO(3–2) moment-1 velocity map with the moment-0 overlaid in black contours. The critical curve from the \cite{Foo_2025} lens model is plotted in magenta (of which ID12 was previously identified as the quadruply imaged system). We highlight the offset between the merging stellar pair, 6.1/6.2 (in cyan circles) and the molecular gas (6.1CO/6.2CO). This CO emission resides on the caustic providing a highly magnified view of these merging cold gas reservoirs (6.1COa/b and 6.2COa/b).}
\label{fig:A1}
\end{figure*}

\subsection{Resolving the Stellar Build-up of DSFG Members} \label{Resolving the Stellar Build-up of DSFG Members}

Among disk galaxies the NIR observations reveal spiral arm structure (ID9, ID10, ID11), most prominently detected in ID10, potentially hosting a stellar bar at $z>2$ \citep{Guo_2023, Le_Conte_2024, Huang_2023, Smail_2023, Umehata_2025}. It exhibits a compact faint dust component limited to its core with relatively low stellar reddening F277W$-$F444W$\sim 0.5$, where these galaxies likely belong to a less dust/gas-rich and more moderate star-forming population representing the fainter end of the IR luminosity function \citep{Franco_2018}. The lower CO(3-2) line emission and dust continuum suggests lower overall gas fractions in these disks that could facilitate the formation of spiral arm and stellar bar structure \citep{Seo_2019}. Additionally, these galaxies reside in the outskirts of the core structure (corrected in the source-plane) relative to the more central members, where environmental effects may not be as strong (i.e., tidal and ram-pressure stripping). It is still unclear how $z>2$ environmental influences may further prohibit or induce the formation and sustainability of (barred-)spirals as compared to local cluster member galaxies \citep{Aquino-Ortiz_2026}.
\par
The most dust-continuum bright and stellar reddened disks (i.e., ID1, ID2), in contrast do not exhibit these more dynamically stable structures (bars, spiral arms). Instead these galaxies exhibit more extended and smooth stellar disk-like morphology with widespread color excess of F277W$-$F444W$\gtrsim 1.5$ and dust continuum emission (See Figure \ref{fig:3}) indicating ongoing, galaxy-wide obscured star formation. In a few galaxies evidence of clumpy structure is resolved in the F277W$-$F444W color maps and specifically in the ALMA continuum TM1 observations. ID2 is the most striking example of this, where the high resolution TM1 imaging detects multiple $\sim$kpc size dust clumps that are even spatially aligned with clumpy structure seen in the F277W$-$F444W maps. Clump formation within these disks may be driven by disk instabilities in-situ \citep{Dekel_2009, Bournaud_2014}, able to sustain galaxy-wide star formation \citep[i.e.,][]{Kamieneski_2024}, demonstrating that such secular processes alone are capable of driving such starburst activity \citep[e.g.,][]{Gillman_2024, Ren_2025}.


Moreover, several members (ID4, ID7, ID11) host rest-frame UV/optically-detected clumps spatially offset from the dust continuum allowing them to penetrate the patchy ISM, further indicating non-uniform stellar-to-dust mixing. The emergence of these clumps may have also originated both ex-situ and in-situ from minor mergers or gas-rich disk instability \citep[e.g.,][]{Rujopakarn_2019}. We also detect smaller, more point-like clumps which may be candidate (super)-stellar clusters at scales of $\sim10$s of pc enabled by lensing. Remarkably, many of these point-like sources are resolved in the extended diffuse emission of ID3 and ID6 (mergers, see section \ref{Section-Mergers}). The detection of such clumps may be enabled due to more moderate dust attenuation of the diffuse tidal emission. In the merger system of ID6 we visually identified $\gtrsim 17$ point-like clumps embedded in the tidal emission, extending even into the outermost regions. 
These stellar clusters may have been gravitationally ejected or resulted from star formation induced by tidal interactions \citep{Renaud_2013}, where the most massive of which can survive tidal destruction and may be the direct progenitors of globular clusters \citep{Bournaud_2008, Claeyssens_2023}. 

A handful of members exhibit bright compact and even point-source stellar components within their cores that may indicate the presence of bulge-formation, central starburst and even potential AGN (ID3, ID6.1/6.2 and ID8.1). ID 8.1---the southern merger--exhibits a point source component in all NIRCam filters, with potential PSF contamination, suggesting the presence of an AGN \citep{Ortiz_2024} or a bright nuclear star cluster, likely triggered by the merger. Bright nuclear cores with compact spheroidal stellar structures are also detected in ID3, and in both galaxies of the ID6 merger pair which could be central starburst driving (early) bulge formation, and indication of their transition to becoming quiescent galaxies \citep[i.e.,][]{Tadaki_2020,Puschnig_2023, Benton_2024}. Furthermore, the tidal disruption of their molecular gas components (see \ref{Section-Mergers} may further accelerate quenching in these merging systems.


\subsection{DSFG Evolution in Protocluster Cores}
The extent to which protocluster core environments affect DSFG evolution at $z>2$ remains unclear, given multiple mechanisms may operate simultaneously. In J0846, we find evidence of interactions, mergers, and disks with distinct morphological features, along with varied stellar color and dust emission properties indicating a heterogeneous population of DSFGs. This suggests that these members, all harbored within a compact protocluster core ($\lesssim$ 300 kpc projected extent and velocity difference of $\Delta V\approx$ 800 km s$^{-1}$), likely reflect different evolutionary pathways toward quiescent cluster galaxies. Recent NIRCam and ALMA studies of DSFGs in the protoclusters have also identified diversity in their samples. \cite{Umehata_2026} find a wide range of stellar and dust continuum properties among the DSFGs in the SSA22 $z=3.09$ protocluster, but with predominantly disklike morphologies indicating more secular stellar mass build-up \citep{Umehata_2025}. Similarly, DSFGs reported in the Spiderweb protocluster at $z=2.16$ are proposed to be driven by cold gas filamentary accretion onto their disks  \citep{Zhang_2026}. 
\par
Comparing $z > 2$ protoclusters demands caution, as such structures may represent distinct large-scale formation stages and scenarios. For example the core of the Spiderweb protocluster already hosts massive quiescent galaxies \citep{Naufal_2024}, where the detected disk DSFGs mostly reside in the outskirts and are expected to grow as they migrate towards the center \citep{Zhang_2024}. In contrast, J0846 and other reported DSFG-rich $z>2$ cores appear to occupy more spatially compact environments \citep{Miller_2018, Wang_2016, Oteo_2018, Champagne_2021} that likely harbor elevated merger and interaction rates. Despite this, we find that disklike properties (i.e, ordered gas velocity gradients, galactic bars and spiral arms) do persist in such environments \citep[e.g.,][]{Venkateshwaran_2024, Kaasinen_2026}, and perhaps can survive even major merger events \citep[i.e., ID4, ID6, e.g.,][]{Zeng_2024}. Furthermore, members with the most wide-spread clumpy dust continuum and reddest stellar colors are disk galaxies (ID1, ID2) suggesting such secular mechanism can drive starburst activity. However, given the proximity, many of the members will likely undergo some merger event if not already interacting.
\par
Evidence of bulge features in the identified merger systems (ID3, ID6, ID8) suggest accelerated evolution towards quiescent elliptical galaxies, highlighting that such events may play a critical role in the quenching and stellar build-up of DSFGs \citep{Araya-Araya_2026} in protocluster cores. J0846 could represent a scenario of the rapid formation of a massive BCG via multiple mergers in the core environment. SPT2349–56, another DSFG-rich protocluster core at $z=4.3$ \citep{Miller_2018, Hill_2020, Zhou_2025}, and potential higher redshift analog to J0846, is predicted to reach final coalescence forming a proto-BCG with a stellar mass of $\textup{ M}_*\sim10^{12} \textup{ M}_\odot$ within $100-300$ Myr \citep{Rennehan_2020, Sulzenauer_2026}.

\par

\par

\section{Summary} \label{Sec:Summary}\

In this paper, we present JWST/NIRCam imaging and ALMA dust continuum observations of J0846+15 `The Koi Pond', a strongly lensed DSFG-rich protocluster core at $z=2.67$. This data provides the first look at the resolved rest-frame NIR stellar and dust continuum properties $>11$ DSFG protocluster members. Our main results are summarized as follows:

\begin{itemize}
\item NIRCam imaging and ALMA band 6 observations successfully detect all 18 CO(3–2) confirmed line emitters reported in \cite{Foo_2025}. The spread in NIR stellar continuum (F444W$=20-23$ mag) and color excess (F277W$-$F444W$=0.5-1.5$ mag) measurements, indicates a range of stellar masses of at least $>1$ dex and dust attenuation reddening of $\Delta$A$_{V}$ $>1$ mag after after comparing to the DSFG sample in COSMOS at similar redshifts with extensive multi-wavelength template fitting estimates.

\item ALMA 1.3mm flux density measurements similarly vary by an order of magnitude $S_{1.3\textup{mm}}\simeq0.1 - 1.9$, revealing relatively dust continuum faint sources of DSFGs in the sample suggesting different levels of ongoing dust obscured star formation. Although aggregate dust continuum versus F277W$-$F444W measurements do not show any correlation, resolved maps suggest the dust surface brightness traces the stellar color excess indicating that the redder colors are driven by dust attenuation as opposed to older stellar populations. However, resolved maps exhibit a wide range of properties including UV/optically-bright emission, both extended and highly compact dust emission indicating different stellar-to-dust geometry among the members.

\item We identify new stellar components and evidence of mergers and interactions in several members (ID3, ID4, ID6, ID8). Prominent extended tidal tails and debris features are detected in NIRCam imaging suggests different merger/interaction stages and scenarios among these systems. Strikingly the CO(3–2) emission of the ID6 is offset by $\sim0.8^{\prime\prime}$ ($\sim2.5$ kpc, magnification corrected) to both stellar merging counterparts. The displacement of this cold molecular gas may be driven by strong tidal forces from the merger and/or tentative evidence of ram-pressure stripping in this protocluster environment. However, CO kinematics suggest that they originate from their progenitor disks that have retained their gas rotation. 

\item Among disk DSFGs in J0846, we find evidence for a myriad of different morphological features including potential galactic bars, spiral arms, nuclear/bulge components and clumpy dust structure. Notably, we find that the DSFGs with clumpy disk structure exhibit the most spatially extended dust emission and wide-spread stellar reddening with clumps spatially aligned in both dust continuum and F277W$-$F444W stellar color maps. This would suggest that their clumpy (obscured) star formation is not restricted to the nucleus but rather galaxy-wide and that such secular processes can trigger starburst events without major mergers. This is in contrast to the less dusty, dynamically stable disks hosting more distinct galactic structures (i.e., bars, spiral arms), likely reflecting different molecular gas supply levels and thus evolutionary stages. 

\item The diverse range of stellar and dust properties across this DSFG-rich protocluster core suggest such environments can host heterogeneous populations, with individual members following divergent evolutionary pathways. However, given the compact structure of J0846, galaxy interaction and merger events likely play an inevitable role in driving their transition towards massive, quiescent ellipticals expected to reside in evolved cluster cores. The identification of multiple mergers hosting prominent bulge-like components provides early signatures of such accelerated evolution.


\end{itemize}

The J0846 \textit{Koi Pond} protocluster offers a rare magnified glimpse into the rapid stellar build-up of large-scale structures. Future work will address differential lensing corrections to recover the intrinsic kinematic and morphological properties of each member. Combined with resolved stellar mass and molecular gas analyses, will allow us to investigate how different physical processes govern stellar mass build-up and shape the eventual descendant cluster cores. 

\par
\par

\acknowledgments
N.F. was supported in part by the Arizona NASA Space Grant Consortium, Cooperative agreement 80NSSC20M0041 and Vivian Forde Graduate Fellowship. N.F. and K.H. would like to thank the European Southern Observatory Office for Science in both Vitacura, Chile and Garching, Germany
for the Science Support Discretionary Fund that enabled N.F. to
visit Chile and focus on this project. K.H. and N.F. would like to
thank Manuel Aravena, Jorge Gonzalez-Lopez, Manuel Solimano,
and others at the Universidad Diego Portales in Santiago, Chile for
useful discussions and perspectives. N.F. would like to thank the
National Radio Astronomy Obsevatory (NRAO) for funding
through the Student Observing Support program. MK acknowledges support from the Australian Research Council via the Discovery Early Career Researcher Award DE250100709. PSK acknowledges financial support from the Knut and Alice Wallenberg Foundation. RAW acknowledges support from NASA JWST Interdisciplinary Scientist grants NAG5-12460, NNX14AN10G and 80NSSC18K0200 from GSFC. AWSM and JB acknowledge the support of the Canadian Space Agency (CSA) [25JWGO4B08] and the Natural Sciences and Engineering Research Council of Canada (NSERC) through grant reference numbers RGPIN-2021-03046 and RGPIN-2026-07024. E.F.-J.A. acknowledge support from UNAM- PAPIIT project  IA104725. We thank Jed McKinney and the collaboration for their help in providing catalogs for the SCUBADive sources.

\bibliography{J08/J08_ref}{}
\bibliographystyle{aasjournal}




\end{document}